\documentclass[12pt,twocolumn]{article}
\usepackage{graphicx}
\usepackage{color}
\usepackage{hyperref}
\usepackage{caption}
\usepackage{setspace}
\usepackage[margin=1in]{geometry}
\usepackage{ulem}
\usepackage[T1]{fontenc}
\usepackage{amsmath}
\usepackage[utf8]{inputenc}
\usepackage{hyperref}
\usepackage{epstopdf}
\onecolumn
\begin{document}
\begin{center}
{\Large {\bf Simulation of Kosterlitz-Thouless (KT) Transition with Classical Monte-Carlo Simulation}}
\end{center}
\vskip 2cm
\begin{center}
{\it Nepal Banerjee}\\
{\it Department of Physics,University of Seoul,South Korea}\\
\vskip 1cm
{\it \textbf{Email:} nb.uos1989@gmail.com}
\end{center}

\vskip 1cm
\begin{abstract}
Spontaneous symmetry breaking of 2D isotropic Heisenberg magnet is restricted by Mermin-Wagner theorem at any finite temperature in presence of short-range exchange interaction.Kosterlitz and Thouless using XY spin model showed that how an order state could developed in 2D spin system in presence of short range isotropic interaction.Very recent discovery of several van der waals magnet revised and redefined our understanding on 2D Heisenberg magnet and its ground state properties.After a rigorous and careful study of several 2D magnetic material we have realized from both experimentally and numerically that the finite size of a 2D system has great impact on the ground state symmetry breaking.Because of that finite size effect more often an anisotropic residual magnetic moment is generated and trigger the spontaneous symmetry breaking at finite temperature (T) and even only presence of short-range interaction we observed the phase transition of that Heisenberg spin system.In this present work we have shown the basic role of finite size,anisotropy during the symmetry breaking of 2D Heisenberg XY magnet.Here we have simulated Kosterlitz-Thouless transition using classical Monte-carlo simulation and study the effect of anisotropy during the phase transition.We presented the behaviour of different thermodynamic properties of 2D XY  spin model system during the Kosterlitz-Thouless (KT) transition.The generic characteristic of KT transition which make it distinct from other critical phenomena is that the peak of heat capacity is not diverging with increase of system size rather peak is decreasing with the increasing of system size near at transition temperature.Here we are observing that behaviour in our present simulation and that specific behaviour help us for classifying the present transition as Kosterlitz-Thouless (KT) transition.    
\end{abstract}
\section{Introduction}
The realization of spontaneous magnetization in two dimension with Heisenberg spin is one of the most challenging research from last several decades and tremendous effort is devoted for realizing this order state\cite{fisher,cardy,kardar,suzuki,tobo,philip,burch2018magnetism,jeil,jeilprl, gong2017discovery,park2016opportunities,jiang2021recent,chit2,
cri3_1,sachdev1999quantum,lado,auerbach}.Mermin-Wagner theorem restrict the realization of long range order(LRO) state for Heisenberg spin in two dimension\cite{Mermin_wagner}.Despite the rigorous proof of this theorem Kosterlitz and Thouless make it possible and able to realize a different type of order state in two dimension in presence of short-range interaction with XY spin\cite{kosterlitz1,kosterlitz2,MA,Hasen}.The order state at low temperature is different from conventional long-range order(LRO).A quasi long-range order(QLRO) state appears and its transition and critical properties is completely different from first order and second order phase transition.Kosterlitz and Thouless has shown that how topological defect can trigger this type of phase transition\cite{kosterlitz2018topological}.We can estimate the transition temperature from simple energy-entropy argument.We can write the Hamiltonian of the present system as 
\begin{eqnarray}
H=-J\sum_{<i,j>} S_i.S_j= -J\sum_{<i,j>} \cos(\theta_{i}-\theta_{j})
\end{eqnarray}
In a harmonic approximation we can express this Hamiltonian as 
\begin{eqnarray}
H &=&-J\sum_{<i,j>} \cos(\theta_i -\theta_j)\\
  &=&-J\sum_{<i,j>} [1-(\theta_i -\theta_j)^2/2 + O((\theta_i -\theta_j)^4)]
\end{eqnarray}
If we consider only nearest neighbour interaction strength J then the energy cost due to the spins at a distance r from the center of the vortex is,in the harmonic approximation,$J/2(2\pi/2\pi r)^2 (2\pi r)$.Therefore the total contribution from a vortex in a system of size L is given by 
\begin{eqnarray}
E_v &=& E_0 + \frac{J}{2}\int_{a}^L |\nabla \theta|^2 d^2r
=E_0 +\frac{J}{2}\int_{a}^L [\frac{2\pi}{r}]dr =E_0 + J \pi ln[\frac{L}{a}]\\
E &=& E_v-E_0
=\frac{J}{2}\int_{a}^L [\frac{2\pi}{r}]dr =J \pi ln[\frac{L}{a}]
\end{eqnarray}
Where "\textbf{a}" is the lattice spacing.We can placed the vortex at any position in a lattice of size ($L\times L$).So the number of way we can placed the vortex is ($L/a\times L/a$).So the entropy of the vortex is 
\begin{eqnarray}
S=Kln(\Omega)=K ln(L/a)^2
\end{eqnarray}
So the free energy of that vortex is 
\begin{eqnarray}
F&=&E -TS =J \pi ln[\frac{L}{a}] -T K ln(L/a)^2\\
&=&(J\pi - 2KT)ln(L/a)
\end{eqnarray}
Now when free energy of vortex F>0 then vortex formation is not allowed in the system and if the free energy F<0 then the vortex formation is preferable and proliferation of vortex destroy the spin ordering.We can estimate that critical temperature at which system goes through a QLRO state to completely disorder state.If we set F=0 then we can estimate the critical temperature as $ T_c=J\pi/2K $ where this transition takes place.So at $T>T_c$ the free energy F<0 and because of proliferation of vortices the quasi long range order is completely destroyed and we observe disorder phase which is characterized by exponential decay of spin-spin correlation.In case of $T<T_c$ the free energy F >0 and vortex formation is not allowed and we observed quasi long range order (QLRO) in our system which is characterized by algebric decay of spin-spin correlation\cite{peter}.This simple argument gives us a qualitative idea about Kosterliz-Thouless transition and we able to estimate the transition temperature($T_c$) based on that argument.This noble and breakthrough work spark the hope of realization of the spontaneous order state in two dimension\cite{chalker1,chalker2,chalker3,hasen2}.Recent discovery of several van der waals magnet shows intrinsic magnetic order state with Heisenberg type of spin\cite{nov1,nov2}.The chiral magnet and skyrmionic phase is another kind of interesting phase in the direction of 2D magnet and topological defect plays a vital role for originating this type of magnetic phase.The emerging properties of skyrmionic phase is deeply connected with the behaviour of XY spin\cite{mol,randeria,sujay}.This kind of low dimensional magnet host several interesting phase and the magnetic order which appears due to the spin-spin interaction has high impact on spin transportation and information storage in ultra-thin spintronics devices\cite{kim2021transport,das2021xy}.The recent discovery of several dissipation less spin current which appears in QSH and QAH phenomena is deeply inter-connected with bulk spin ordering and dynamics\cite{nomura2006quantum,nepal_topo_2,sinova2015spin,sinova2004universal,nagaosa2010anomalous,gao2021layer,
bernevig2022progress,kubler2014non}.It is still not very clear and unknown that how bulk spin dynamics of 2D XY spin impact on edge spin current in two dimensional ultra-thin ribbon of magnetic materials.Recently discovered moire magnet also a rich platform and hosting plethora of correlated phase and still it is unknown how order state is developing in this type of moire magnet for 2D XY spin and what is the role of interlayer spin-spin interaction in stabilizing the spontaneous magnetic ordering in this moire magnet.So there are several challenging question which are still unknown and that prepare the study of 2D XY spin system more exciting research field in the contest of moire magnet and van der waals magnetic material\cite{bistritzer2011moire,lu2019superconductors,qiao2011electronic,hejazi2020noncollinear}.Also we know that the phase transition of 2D XY model and bosonic superfluid follows the same universality class.So after realizing  all those challenging questions and its impact we are trying to explore and simulating the Kosterlitz-Thouless (KT) transition using 2D XY spin and study the  critical phenomena for simple square lattice with classical Monte-Carlo simulation.Here we are trying to  get some idea of spins behaviours from microscopic length scale.We first study the isotropic 2D XY spin model where we consider ferromagnetic exchange interaction where J>0.Here for simplicity we have considered J=1.Next we have studied the effect of anisotropy($\Delta$) in 2D XY spin model system.We have simulated the critical phenomena for both isotropic and anisotropic case.We observed the spin dynamics of 2D XY spins with temperature(T) and observe how order state is developing here at lower temperature.We have organized our brief paper in following way.First we have described the model Hamiltonian in section-2 and in section-3 we have described the simulation methodology and results.In last section we have discussed about the results and finally make a conclusion.    
\section{Model Hamiltonian}
\subsection{2D Isotropic XY Model:}
Here we have considered 2D XY spin model Hamiltonian and consider planner type of spins.We have considered our model Hamiltonian as 
\begin{eqnarray}
H=-J\sum_{<i,j>} (S_i^x S_j ^x + S_i ^y S_j ^y)
\end{eqnarray}
We have considered $S=(S_x,S_y)$.Here we have defined $S_x=|S|cos(\phi)$ and $S_y=|S|sin(\phi)$.Here we consider two dimensional planner type of spin which we says as XY type of spin.Here we have defined the two component $S_x=S_x(i,j)$ and $S_y=S_y(i,j)$.Here we have consider J=1 and |S|=1.
\subsection{2D Anisotropic XY Model:} 
 Here we have consider 2D XY spin model Hamiltonian and consider planner type of spins.We have consider our model Hamiltonian as 
\begin{eqnarray}
H=-\sum_{<i,j>} (J_x -\Delta) S_i^x S_j ^x + J_y S_i ^y S_j ^y 
\end{eqnarray}
We have considered $S=(S_x,S_y)$.Here we have defined $S_x=|S|cos(\phi)$ and $S_y=|S|sin(\phi)$.Here we consider two dimensional planner type of spin which we says as XY type of spin.Here we have defined the two component $S_x=S_x(i,j)$ and $S_y=S_y(i,j)$.Here we have consider $J_x=J_y=1$ and |S|=1.

\section{Simulation Methodology and Results:}

\subsection{2D Isotropic XY Model:}
Here we will going to describe the methodology and results of our simulation\cite{banerjee2023critical,drouin2022kosterlitz}.Here we have been focusing ourself strictly on isotropic XY Heisenberg spin model and this spin are simulated in 2D square lattice grid after choosing an angle $\phi$ randomly such a way that the random angle $\phi$ must be uniformly distributed between $0$ to $2\pi$.We have chosen random $\phi$ as $\phi=2\pi u_1$.Here $u_1$ is a computer generated random number and which is a tested as uniform random number generator.We have assigned the spin components of that XY spin as $S_x(i,j)=|S|cos(\phi)$ and $S_y(i,j)=|S|sin(\phi)$ in 2D square lattice grid over unit circle,where $S_x,S_y$ are two component of that spin.The way we have chosen this components such that the  spin magnitude must be normalized to the value of |S|.Each spin has its independent coordinate (i,j) in that 2D square lattice grid.In this way we have simulated random spin configuration where all the spins has randomly chosen its $S_x$ and$S_y$  component and which are sitting on independent 2D square lattice  grid (i,j).We have assigned this random configuration of spins as a high temperature spin configuration,which is far above the transition temperature.Then we have started to cooling the system and decreasing the temperature with small steps.In each step we have measured and calculate the different thermodynamic quantity like spontaneous magnetization,susceptibility,heat-capacity,energy.We have measure all those thermodynamic quantity after taking ensembles average with $3\times10^4$ number of equivalent spin configuration at a particular temperature.We here measure spontaneous magnetization which is square root average of magnetization of each component of a spin and we write it as a $M=\sqrt{(m_x^2 +m_y^2 )}$.Here $m_x$,$m_y$ are the average spin component along the $S_x,S_y$ direction respectively for each site of the lattice grid.At each temperature we take the average over $3\times10^4$ identical spin configuration which we call as ensembles for that particular temperature.We ignore $3\times 10^4$ ensembles just to reach equilibrium.Here we have calculated susceptibility,heat-capacity at each temperature using the same procedure mentioned before.We use the following formula for that calculation of susceptibility,which is $\chi=L^2(<M^2>-<M>^2)/T$ and for heat-capacity we are using the following formula $C_V=L^2(<E^2>-<E>^2)/T^2$. Here the energy has continuously decrease with temperature and it indicating that the system is going towards the reaching  of stable equilibrium state which is here ferromagnetic ground state.
\begin{figure}[htp]
\centering
\includegraphics[scale=0.380]{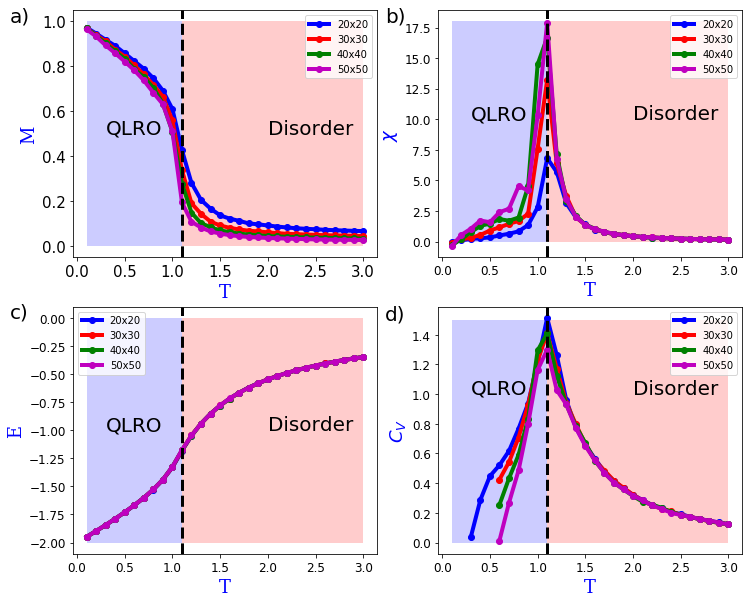}
\caption{Here we have presented the different thermodynamic variables with T for isotropic XY model.Here J=1 and |S|=1.We have presented the study for $20 \times 20 ,30 \times 30 ,40\times 40 ,50\times 50 $ lattice size. a) Here we have presented spontaneous magnetization(M) with T.b)Here we have presented susceptibility($\chi$) with T. c)Here we have presented the behaviour of energy(E) with T.d) Here we have presented the behaviour of heat-capacity($C_v$) with T.Here we observe a transition at $T_{CV}=1.1$ which we mark with a vertical dotted line and $T_{CV}>T_{BKT}$.Here we are observing a peak at $C_v$ rather than singular behaviour and that peak at $C_v$ is decreasing with increasing of system size and that specific features help us to identify this phase transition as KT transition.}
\label{}
\end{figure}
Here we have observed the peak of susceptibility and heat-capacity near at T=1.0.The generic characteristic of KT transition which make it different from other universality class is revealed in the properties of $C_v$.
\begin{figure}[htp]
\centering
\includegraphics[scale=0.30]{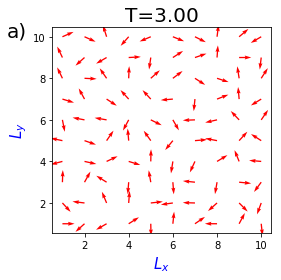}
\includegraphics[scale=0.30]{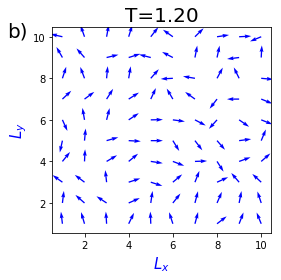}
\includegraphics[scale=0.30]{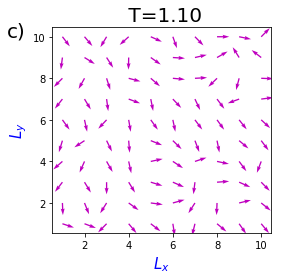}
\includegraphics[scale=0.30]{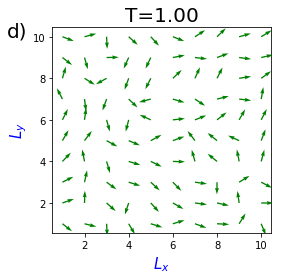}
\includegraphics[scale=0.30]{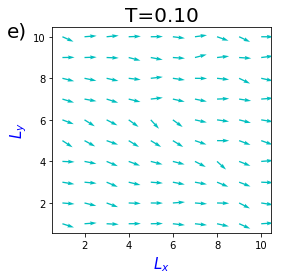}

\caption{Here we have represented the spin texture of isotropic XY model for |S|=1 and J=1 for $10 \times 10 $ lattice at different temperature.a) Here we have presented the spin texture at T=3.00 which is far above transition temperature and consider as high temperature configuration.Here we observed that the spins are completely disorder and have no correlation with each other. b)Here we have presented the spin behaviour for T=1.2 which is closed to transition point T=1.1.Here we can observe  short range correlation between spins and the dynamics are showing correlated behaviour.c)Here we presented the spin texture for T=1.1 where the transition takes place.Here we can visualize spin vortex and anti-vortex.d)Here we have presented the spin texture for T=1.0 and spins are showing some ordering and correlated behaviour.e) Here we have presented the spin texture for T=0.1 and we are observing a QLRO in our spin system.}
\label{}
\end{figure}
Here the peak of heat capacity is decreasing with increase of system size(L) and we observed this phenomena in our present simulation results.Here we have studied the whole simulation for $20 \times 20, 30 \times 30 ,40 \times 40,50 \times 50  $ lattice site which is basically a square lattice grid.Here for a particular temperature we have equilibrate the spin configuration with $6\times 10^4$ MC steps among that we discard $3\times 10^4$ just to achieve equilibrium and consider $3\times 10^4$ steps for average after achieve that equilibrium spin configuration.In each step we use single spin flipping  Metropolis algorithm to equilibrate the spin configuration.Here we are going to discuss about that simulation methodology in details.Here we have selected randomly a site,let suppose it is (i,j) point.Then we have calculated the energy of that spin because of nearest neighbour interaction with its neighbouring spin.Now we have created a new direction of spin state after choosing a random $\phi$ angle using the formula $\phi=2\pi u_1 $.Here $u_1$ is a computer generated  random number,which we have already tested as a uniform random number generator.We have flipped the selected spin along that newly created random $(\phi )$ direction.Now we have calculated the energy of that flipped spin based on the interaction with its neighbour spin.Let suppose that before flipping the energy of that spin was E and after flipping the energy of that spin become $E_1$.Now calculate the energy difference $\Delta=(E_1 -E)$ and using that difference we have calculated the flipping  probability using the formula $P=e^{-\Delta/T}$.If that flipping probability P is greater then the random number then we have flipped that selected spin along the new direction otherwise we have kept that selected spin in the previous state.We have selected another spin and repeat that same procedure and we have continue this selection process of spin site up to $L \times L $ times so that each site get atleast a chance of being selected randomly.We call this method Metropolis flipping and we have done this process $6\times10^4$ times.Basically with this Metropolis flipping the spin configuration start to equilibrate slowly and create equilibrium unique spin configuration which we call as ensemble at a particular temperature and those equilibrium spin configuration is only consider for taking average of different thermodynamic variables.So finally we have set $3\times 10^4$ steps just to equilibrate the system using Metropolis flipping.Once the system reach at equilibrium then we have calculated different thermodynamic quantity using ensemble average.Here for taking average we use $3 \times 10^4$ ensembles.Here we have calculate the average value of $m_x$,$m_y$ with the formula $m_x=\sum S_x/L^2$ and $m_y=\sum S_y/L^2$.In this calculation we have calculated magnetization (m) using the formula $M=\sqrt{m_x^2 +m_y^2}$.We observe that magnetization is change from $10^{-2}$ oder to 1  continuously when temperature(T) cool down from a higher temperature to lower temperature via the transition temperature $T=T_c$.
\begin{figure}[htp]
\centering
\includegraphics[scale=.300]{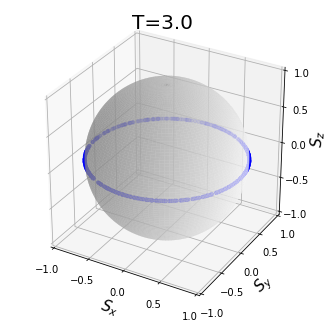}
\includegraphics[scale=.300]{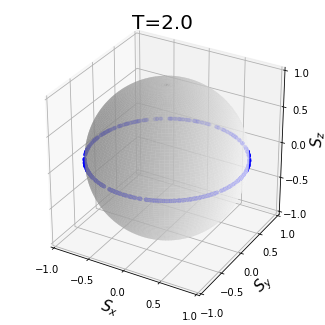}
\includegraphics[scale=.300]{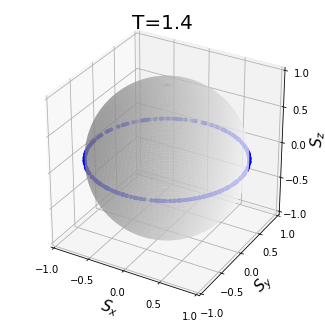}
\includegraphics[scale=.300]{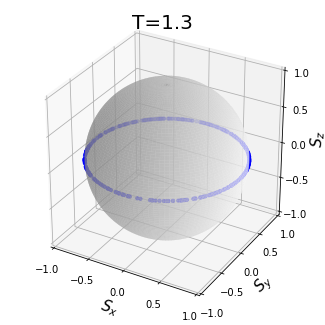}
\includegraphics[scale=.300]{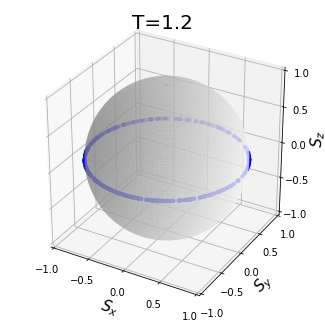}
\includegraphics[scale=.300]{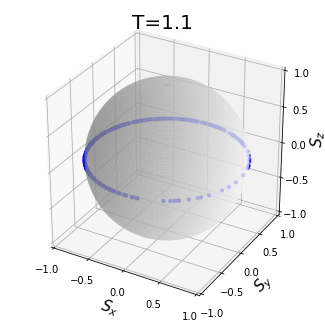}
\includegraphics[scale=.300]{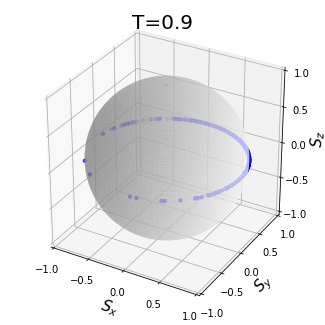}
\includegraphics[scale=.300]{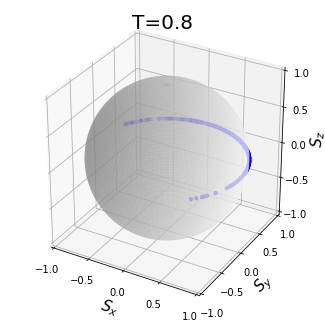}
\includegraphics[scale=.300]{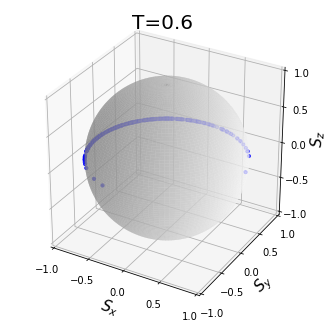}
\includegraphics[scale=.300]{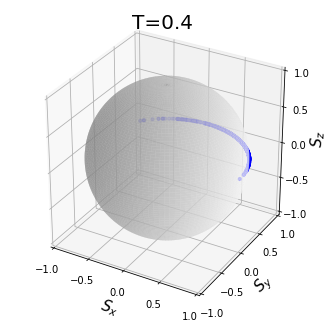}
\includegraphics[scale=.300]{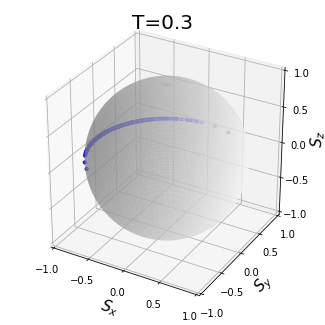}
\includegraphics[scale=.300]{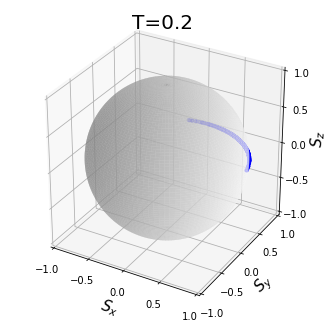}
\caption{Here we have presented the spins configuration which are spin component projection on the surface of a unit sphere which have radius |S|=1 in orthogonal coordinate $S_x-S_y-S_z$.Since it is XY spin so at high temperature spin has O(2) symmetry and spin component projection are uniformly distributed on equator which is presented at T=3.0.When we cooling down our system  we are observing symmetry breaking of the spin configuration which we have presented at different T.Here we have studied this spin dynamics for $20 \times 20 $ lattice system and here |S|=1 and |J|=1.}
\label{}
\end{figure}
We observed  the final stable equilibrium spin configuration for which the  spontaneous magnetization is settled down to 1 finally.
The variation of spontaneous magnetization here sharply indicate a phase transition from disorder to order state and system goes through a continuous transition.The value of spontaneous magnetization change in the oder of $\sim 10^{-2}$ to 1.
We have also observed how the magnetization peak changes with different system size.We have calculated average of $M^2$ which is denoted as $<M^2>$ and average of magnetization which we call $<M>$ at a particular T.This last two quantity is needed for calculate the susceptibility using the formula
\begin{figure}[htp]
\centering
\includegraphics[scale=0.300]{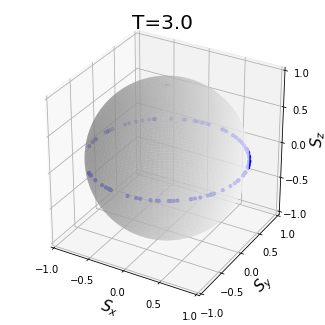}
\includegraphics[scale=.300]{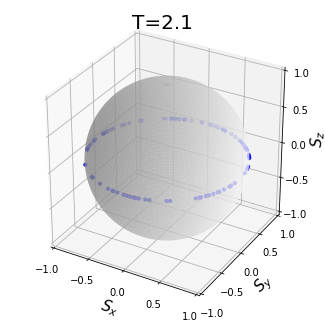}
\includegraphics[scale=.300]{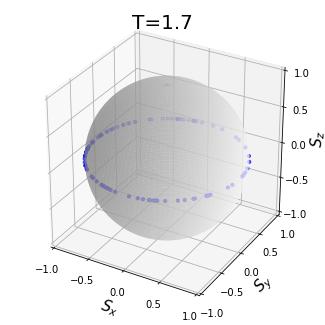}
\includegraphics[scale=.300]{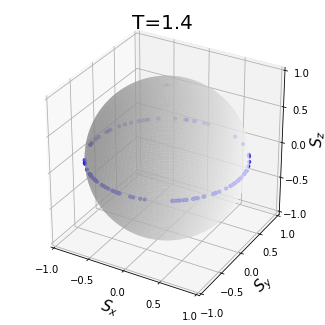}
\includegraphics[scale=.300]{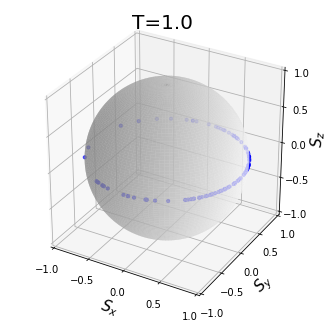}
\includegraphics[scale=.300]{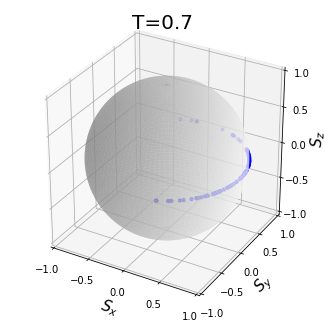}
\includegraphics[scale=.300]{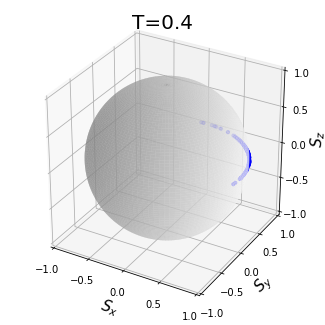}
\includegraphics[scale=.300]{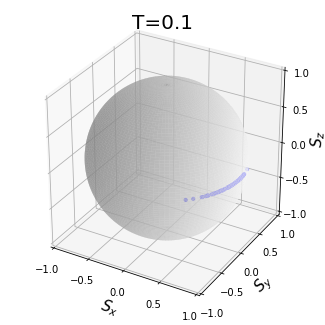}

\caption{Here we have presented the spins configuration which are spin component projection on the surface of a unit sphere which have radius |S|=1 in orthogonal coordinate $S_x-S_y-S_z$.Since it is XY spin so at high temperature spin has O(2) symmetry and spin component projection are uniformly distributed on equator which is presented at T=3.0.At different T  we have presented the symmetry breaking of the spin system when we cooling down our system.Here we have studied this spin dynamics for $10 \times 10 $ lattice system and here |S|=1 and |J|=1.}
\label{}
\end{figure}
\begin{eqnarray}
\chi=L^2(<M^2>-<M>^2)/T
\end{eqnarray} 
Similarly we have calculated the average value of energy of  each spin which we denote as $<E>$ and square of energy $<E^2>$. After that we have calculated the heat-capacity using the following formula
\begin{eqnarray}
C_V=L^2(<E^2>-<E>^2)/T^2
\end{eqnarray}   
Here the $\chi$ measure the fluctuation of magnetization which is diverge during the  phase transition.Because of that reason we observe a susceptibility peak at Tc.This peak increase with system size and we can observed it (Fig:1) from our calculation very easily.This kind of behaviour easily gives us the idea of the system's behaviour in thermodynamic limit. But here we observed completely opposite behaviour for heat capacity.Here heat capacity peak decrease with system size and that is the generic properties of KT transition.
\subsection{2D Anisotropic XY Model:}
Here we have observed the behaviour of 2D XY model in presence of anisotropy($\Delta$).We observed the effect of anisotropy for the lattice size $10 \times 10 $.We have studied the behaviour of anisotropy for the Hamiltonian in Equation(2).Here we have observed the critical behaviour for anisotropy strength($\Delta$)=0.0-0.6 for a $10 \times 10$ lattice size.We have observed that with increasing the strength of anisotropy $\Delta$ the transition temperature of that system keep on increasing.We have followed the same methodology for the simulation as we described in previous section.Here we have notice the behaviour of the XY spin texture in XY plane and the spin component projection on unit sphere in $S_x-S_y-S_z$ coordinate.Here we have considered a unit sphere and since our system has O(2) symmetry so all the points on equator indicates a distinct spin state  which are sitting at square lattice grid.That distribution is uniform at high temperature and we can visualize that uniform distribution from spin component projection at high temperature.
\begin{figure}[htp]
\centering
\includegraphics[scale=0.30]{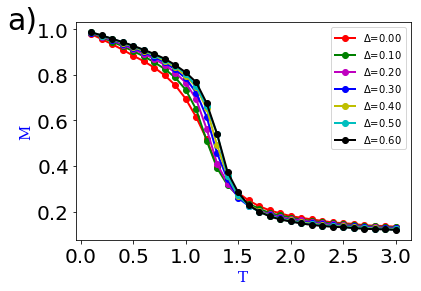}
\includegraphics[scale=0.30]{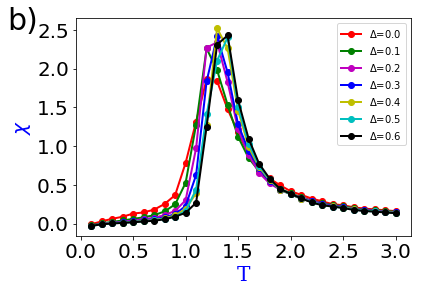}
\includegraphics[scale=0.30]{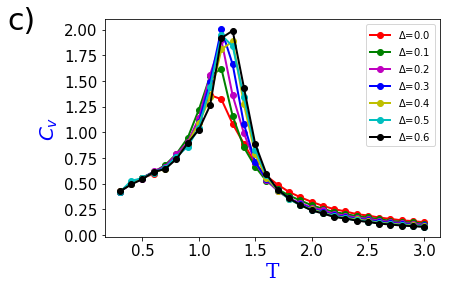}
\caption{Here we have presented the different thermodynamic variables for $10 \times 10 $ lattice size with anisotropy($\Delta$) range from 0.0-0.6.Here we have done this study for $J_x=J_y=1$ and |S|=1.a)This represent the behaviour of spontaneous magnetization (M) with T for anisotropy($\Delta$)=0.0-0.6.b)This represent the variation of susceptibility($\chi$) with T for anisotropy($\Delta$)=0.0-0.6.c)This represent the behaviour of $C_v$ with T for anisotropy ($\Delta$)=0.0 -0.6.Here we are observing peak at $\chi$ and $C_v$ at transition point and transition temperature and that peak is increasing with $\Delta$ which is clear from $C_v$ vs T plot.}
\label{}
\end{figure}
So  we notice that at high temperature all the spin components reside on equator uniformly.When we are cooling down our system then all the spins start to show correlated behaviour and that behaviour is more distinct near at transition temperature.We observed QLRO state in our spin system  at low temperature (T=0.1) and finally all the spins are settled in a particular direction according to the strength of exchange interaction.Here since we choice our exchange interaction in a particular fashion where  $s_x-s_x$ interaction $J_x$ decrease with the increment of $\Delta$ and all the spin will prefer to orient themselves along the $S_y$ direction.We have also observed that the behaviour of spin texture in spatial $L_X-L_Y$ plane.We notice that ground state symmetry breaking of spin with anisotropy($\Delta$).When anisotropy ($\Delta$) is set to zero then the  all spins are oriented along the diagonal of $S_x-S_y$ plane which is actually  equatorial plane of that unit sphere in $S_x-S_y-S_z$ coordinate.We have observed the spin dynamics with temperature.We observe how all the spins are settling in the ground state at low temperature for a particular combination of $J_x$ and $J_y$ in presence of anisotropy($\Delta$).When we keep on increasing $\Delta$ then the spin are finding a bias direction to align and all the spins are keep on aligning along $S_y$ direction which is clearly observed from the spin projection on unit sphere in $S_x-S_y-S_z$ coordinate. 
 Here we use the formula for magnetization as 
 \begin{eqnarray}
 M=\sqrt{m_x^2 +m_y^2}
 \end{eqnarray}
 Here $m_x=\sum S_x/L^2$ and $m_y=\sum S_y/L^2$. Here we have used the formula for estimate the susceptibility as 
\begin{eqnarray}
\chi= L^2(<M^2>-<M>^2)
\end{eqnarray}
Here $<M^2>$ is average of $M^2$ in a particular temperature and <M> is the average of M in a particular temperature.
Similarly we have calculated heat-capacity with the formula as 
\begin{eqnarray}
C_V=L^2(<E^2>-<E>^2)
\end{eqnarray}
Here $<E^2>$ is the average of $E^2$ and <E> is the average of E at particular T.We have take the ensemble average using number of equivalent spin configuration at particular T after reaching the thermodynamic equilibrium.Here we have observed that both of the response function $\chi$ and $C_V$ shows singular behaviour near at transition temperature and peak of $C_v$ increase with $\Delta$. 
\section{Discussion and Conclusion:}
Here we have studied successfully the critical phenomena of 2D XY model in square lattice and correctly predict the transition temperature($T_c$).We have observed the dynamics of planner spins during the transition at different temperature.We observed the microscopic picture of spin dynamics and how symmetry breaking is happening during the transition.We have notice that the exchange interaction $J_x$ and $J_y$ has very vital role in spontaneous symmetry breaking.Here we have studied the properties of quasi long range order (QLRO) in our spin system and here we have observed the microscopic spin dynamics during the development of QLRO.  
\section{Acknowledgement}
N.B  is greatly acknowledging University of Seoul(UOS) for the funding from SAMSUNG and NRF project at initial stage of this work and we are greatly acknowledging IIT kanpur for giving visiting scholar position and providing generous research facility during the visit.

\end{document}